**Stacking-symmetry governed second harmonic generation in graphene trilayers**

Yuwei Shan,[1] Yingguo Li,[1] Di Huang,[1] Qingjun Tong,[2] Wang Yao,[2] Wei-Tao Liu,[1,3] Shiwei Wu[1,3*]

[1] State Key Laboratory of Surface Physics, Key Laboratory of Micro and Nano Photonic Structures (MOE), and Department of Physics, Fudan University, Shanghai 200433, China.

[2] Department of Physics and Center of Theoretical and Computational Physics, University of Hong Kong, Hong Kong, China.

[3] Collaborative Innovation Center of Advanced Microstructures, Nanjing 210093, China.

[*] Corresponding email: swwu@fudan.edu.cn

**Abstract**

**Crystal symmetry plays a central role in governing a wide range of fundamental physical phenomena. One example is the nonlinear optical second harmonic generation (SHG), which requires inversion symmetry breaking. Here we report a unique stacking-induced SHG in trilayer graphene, whose individual monolayer sheet is centrosymmetric. Depending on layer stacking sequence, we observe a strong optical SHG in Bernal (ABA) stacked non-centrosymmetric trilayer, while it vanishes in rhombohedral (ABC) stacked one which preserves inversion symmetry. This highly contrasting SHG due to the distinct stacking symmetry enables us to map out the ABA and ABC crystal domains in otherwise homogeneous graphene trilayer. The extracted second order nonlinear susceptibility of the ABA trilayer is surprisingly large, comparable to the best known 2D semiconductors enhanced by excitonic resonance. Our results reveal a novel stacking order induced nonlinear optical effect, as well as unleash the opportunity for studying intriguing physical phenomena predicted for stacking-dependent ABA and ABC graphene trilayers.**

**Introduction**

The van der Waals nature of layered crystals can lead to profound effects of stacking order on electronic (*1, 2*), optical (*3-5*), magnetic (*6*), piezoelectric (*7*) and mechanic (*8, 9*) properties. For instance, the van der Waals stacking evokes a drastic change in nonlinear optical response in atomically thin 2D materials (*10-15*). A notable example is semiconducting transition metal dichalcogenides $MX_2$ (M = Mo, W; X = S, Se). Monolayer $MX_2$ lacks the crystal inversion symmetry and thus permits the electric-dipole allowed second harmonic generation (SHG), which can be further enhanced by excitonic effect (*10, 11, 14, 15*). In sharp contrast, SHG vanishes in Bernal stacked bilayers or further enhances in rhombohedral stacked bilayers, due to the destructive or constructive interference between neighboring monolayers (*10, 14, 15*). Because SHG is highly sensitive to the structural symmetry, SHG spectroscopy and microscopy has emerged as a powerful and useful technique to study these atomically thin materials.

Among the family of 2D materials, graphene is the most prominent material and has the simplest atomic structure. Yet different from many 2D materials like hBN, $MoS_2$ and GaSe *(12, 13)*, graphene monolayer is centrosymmetric with the inversion center at the hollow site of hexagon (Fig. 1A). Thus, no SHG is expected under the electric dipole approximation (*16*). When two centrosymmetric objects are brought together along the same orientation, no matter how they are translated with each other, the composite system remains centrosymmetric. This rule readily applies to AB stacked graphene bilayer (Fig. 1B). Therefore, SHG is generally unexpected for graphene bilayers as well unless the interface and curvature effects, or electric-dipole forbidden terms are considered (*16*).

However, crystal symmetry can be readily modified in graphene trilayers. As schematically shown in Figs. 1C and 1D, ABA and ABC graphene trilayers differ on the stacking sequence (*17, 18*). Such a difference result in distinct structural symmetries, of which the former is non-centrosymmetric and the latter is centrosymmetric. Therefore, trilayer graphene would provide a platform for exploring a novel nonlinear optical effect, i.e. the stacking-induced SHG. While such an effect is unique to atomically thin 2D materials, it could be extended to other van der Waals stacked materials with centrosymmetric monolayer such as 1T phase transition metal dichalcogenides (*19*).

Here we report the observation of emergent electric-dipole allowed SHG in graphene trilayer with ABA (Bernal) stacking, but it is forbidden in ABC (rhombohedral) stacking. While such difference in stacking order could be captured in their contrasting SHG response, a simple analysis on structural symmetry does not lead naturally to strong SHG. We surprisingly found that the stacking-induced SHG is exceptionally strong, exceeding that of hBN monolayer (*13*) and even comparable to that of $MoS_2$ monolayer upon excitonic resonances (*12*). The ABA trilayer can thus be an outstanding material for exploring nonlinear optical devices. Furthermore, polarization-resolved SHG measurement unambiguously determines the crystalline orientation of ABA trilayer. Because of the distinct stacking symmetry and SHG response between ABA and ABC trilayers, the SHG microscopy provides a superior method to image the ABA and ABC domains in graphene trilayers (*20, 21*). With the aid of this stacking-induced SHG, we believe many interesting physics predicted for ABA and ABC graphene trilayers could be experimentally explored. These include: flat band high temperature superconductivity (*22*), gate tunable anti-ferromagnetic to ferromagnetic phase transition (*23*), band topology associated excitonic physics (*24*), and topologically protected valley-Hall kink and edge states (*25*), and many others (*26, 27*).

**Results**

Few-layer graphene samples were prepared by mechanical exfoliation from Kish graphite on a 300 nm $SiO_2$ on silicon substrate. Figure 2A shows the bright-field optical microscopy image of a mechanically exfoliated few-layer graphene sample. The layer thickness (monolayer, bilayer, and trilayer) was identified by optical contrast and confirmed by Raman spectra (see Supplementary Fig. S1) (*28*). Upon femtosecond laser irradiation, strong nonlinear optical emission can be detected from graphene. Figure 2B shows the optical emission image of the same area in Fig. 2A, which was excited at 1300 nm with the light emission acquired within 425-675 nm. Figure 2C shows the emission spectra from different graphene layers in Fig. 2B. The emission is mainly due to an up-converted, broadband nonlinear photoluminescence (NPL) signal, which arises from the Auger-like scattering of photoexcited hot carriers, and its intensity increases proportionally to the layer thickness (*29, 30*). Besides the NPL, a sharp peak at 433 nm due to the third harmonic generation (THG) was also observed (*31*). As expected, THG is electric-dipole allowed for all graphene layers, and its intensity grows with the layer thickness.

In addition to these known nonlinear optical features, a new and strong peak at 650 nm, twice energy of the incident photon, emerged from the trilayer (Fig. 2C). To confirm it being second harmonic generation (SHG), we tuned the excitation wavelength and found the signal to be always at the half wavelength. We also measured the power dependence of the signal by subtracting the broad background due to NPL. As plotted in the inset of Fig. 2C, the signal intensity grew quadratically with the excitation power, confirming that it is the SHG from trilayer graphene. In comparison to trilayer, SHG is not observed for the graphene monolayer, and nearly absent for the bilayer (Fig. 2C), consistent with their structural inversion symmetry. The weak SHG from the bilayer is likely due to the breaking of inversion symmetry by the substrate or residual doping (*2*).

As we mentioned, though the graphene monolayer, bilayer, and bulk graphite are all centrosymmetric, the overall inversion symmetry can be broken in trilayers of particular stacking orders. This is the case for the ABA (Bernal) stacked trilayer, which is also the most abundant, naturally existing stacking order. In analogy with a $MoS_2$ monolayer, the ABA graphene trilayer has a 3-fold, instead of 6-fold, rotation axis along the surface normal, and three in-plane 2-fold ($C_2$) axes along arm chair directions, which belongs to the non-centrosymmetric $D_{3h}$ point group (Fig. 1C) (*10*). Therefore, upon normal incidence, when the excitation polarizer and signal analyzer are set parallel, there should be $I_{SHG}^{XX} \propto \left| \chi_{aaa}^{(2)} \cos3\phi \right|^2$, with $\chi_{aaa}^{(2)}$ being the responsible susceptibility tensor element, and $\phi$ the angle between the armchair direction and excitation polarization (*10*). When the two polarizations are set perpendicular, we then have $I_{SHG}^{XY} \propto \left| \chi_{aaa}^{(2)} \sin3\phi \right|^2$. Such 6-fold anisotropies of $I_{SHG}^{XX}$ and $I_{SHG}^{XY}$ are clearly seen in Fig. 3A and Fig. 3B, respectively. In comparison, the NPL remains isotropic with respect to $\phi$, reflecting the incoherent nature of hot carrier scattering (*29, 30*). Therefore, the polarization analysis of SHG on graphene trilayer in Fig. 2A reveals the ABA stacking order and the nature of strong SHG as well. The rotational anisotropy also allowed us to determine the crystal orientation of graphene layers, as sketched in Fig. 2A.

In mechanically exfoliated trilayers, there is a common rhombohedral stacking fault leading to the ABC stacking order. It often forms interconnected domains with ABA trilayer and constitutes only about 15% of the total area (*20*). It is thus difficult to identify the stacking order and different domains noninvasively and accurately. Raman spectroscopy and microscopy is the most widely used technique (*28*). To identify the stacking orders between ABC and ABA, subtle changes in the line shape of various Raman modes (*20, 21*), or frequency shifts of interlayer shear mode at very low wavenumbers (*32-34*) are used. Nonetheless, Raman spectra are often weak, and highly susceptible to defects, disorders and strains in the sample that can obscure spectral features (*28*). Furthermore, it is time consuming to microscopically image the ABA and ABC domains by collecting and analyzing the Raman spectra at each pixel.

Because ABC trilayer is centrosymmetric and has distinct structural symmetry from ABA trilayer, no SHG response is expected for ABC trilayer. Thus, SHG microscopy should be a powerful technique to distinguish the stacking orders. Figure 4A shows the bright-field optical microscopy image of an exfoliated graphene sample on the $SiO_2$/Si substrate containing both bilayer and trilayer. The entire trilayer region has a uniform contrast in reflectivity. However, on the nonlinear optical image excited at 1266 nm, different contrasts show up within the same trilayer (Fig. 4B). The contrast of the trilayer region could become uniform again, if we tune the excitation wavelength to 1300 nm so that the SHG signal is no longer in the spectral window (Supplementary Fig. S2). Figure 4C displays emission spectra from the brighter and dimmer regions, respectively. On top of NPL, a prominent SHG peak at 633 nm appears for the brighter region, but is absent for the dimmer one, showing the latter to have a centrosymmetric stacking order. The NPL of the brighter region is slightly weaker at the presence of a strong SHG, possibly due to the conservation of oscillator strength. Meanwhile, the SHG rotational anisotropic spectra in Figs. 4C and 4D proves that the brighter region has a 3-fold symmetry corresponding to the ABA stacking order, and allows us to further determine the crystalline orientation as overlaid in Fig. 4A. To confirm the dimmer and centrosymmetric region to be ABC stacked, we took Raman spectra as shown in Figs. 4E and 4F. Compared to the ABA trilayer, the dimmer region has a slightly weaker 2D band, with the spectral weight shifting slightly to lower frequency, consistent with that reported for ABC trilayers (*20, 21*). Nonetheless, without prior knowledge, it is very difficult to identify the minority ABC stacking based on the subtle differences of the line shape in Raman spectra. SHG clearly yields a much better contrast between the two polymorphs with a high throughput.

Since the inversion symmetry breaking of ABA trilayers is due to the layer stacking, one might consider it as a secondary effect and assume the SHG to be weak. However, by estimating the second order nonlinear susceptibility described in Ref. 12, we surprisingly found $|\chi^{(2)}_{aaa}| \sim 0.9 \times 10^{-10}$ m/V (excited at 1300 nm), which exceeds that of hBN monolayer ($|\chi^{(2)}_{aaa}| \sim 0.3 \times 10^{-10}$ m/V at 900 nm) (*13*), and is even comparable to that of MoS$_2$ monolayer ($|\chi^{(2)}_{aaa}| \sim 4.3 \times 10^{-10}$ m/V at 1600 nm, and $34 \times 10^{-10}$ m/V at 1336 nm) upon excitonic resonances (*10, 12, 13*). In retrospect, such a large optical nonlinearity in ABA trilayer is greatly benefited from the nonlinear optical transitions in resonance with its semi-metallic band structure (*17, 18*), as illustrated in Fig. S3. Additionally, the large second order nonlinearity is also in accordance with the strong piezoelectric effect previously observed among few-layer graphene samples (*7*), although the distinct symmetry in ABA trilayer was overlooked. If the trilayer is ABC stacked, the strong piezoelectric effect would vanish.

**Discussion**

The observation of stacking symmetry governed SHG in graphene trilayers permitted the experimental exploration of many intriguing physical phenomena for ABA and ABC trilayers (*17, 18, 22-27, 35*). As shown in Supplementary Fig. S4 by following the low-energy effective models (*36, 37*), the calculated band structures on ABA and ABC graphene trilayers have distinct features. The ABA trilayer is a semimetal with gate-tunable band overlap. In contrast, the ABC trilayer is a semiconductor with an electrically tunable bandgap, much like that of the AB stacked bilayer (*2*). While the electric and optical conductivities have been confirmed by transport and infrared spectroscopy measurements (*17, 18, 35*), many more features are yet to explore. For ABA trilayer, the effect of trigonal warping is significantly stronger than those in graphene monolayer and ABC trilayer. With the aid of polarization-resolved SHG to determine the crystalline orientation, the anisotropic response and the role of valley pseudospin (*38*) could be studied. For ABC trilayer, two nearly flat bands across the gate-tunable bandgap are formed, and the very high density of states at the band edges opens an opportunity for investigating many exotic phenomena (*22-25*). The SHG microscopy could help fabricate the high-quality ABC trilayer samples even that they are encapsulated inside hBN (*39*), because the SHG vanishes in even layers of hBN due to Bernal stacking and is weak in odd layers.

**Methods**

**SHG measurement.** The nonlinear SHG measurements were conducted with a custom-built sample scanning confocal optical microscope in a back-scattered geometry. The fundamental beam (120 fs, 80 MHz) was linearly polarized and tunable between 680-1300 nm, and focused by a microscopic objective (100×, 0.95 NA) to the diffraction limit onto the sample at normal incidence. The reflected nonlinear optical signal was collected by the same objective. After passing through a dichroic beamsplitter and short-pass (or band-pass) filters, the fundamental beam was filtered out and the nonlinear optical signal was detected by either a single-photon counting silicon avalanche photodetector, or a fiber-coupled spectrograph equipped with a liquid nitrogen cooled silicon charge-coupled device. Nonlinear optical microscopic images were obtained by raster scanning the sample on a piezo-actuated two-dimensional nanopositioning stage. The typical dwell time at each pixel is ~25 ms, more than two orders of magnitude shorter than that used in Raman spectroscopic imaging. To measure the azimuthal anisotropy pattern of nonlinear optical signal, we rotated the polarization of fundamental beam with respective to the graphene surface normal using an achromatic half-wave Fresnel rhomb inserted between the dichroic beamsplitter and the microscopic objective. Polarization state of the signal beam passing through the same rhomb was analyzed by a linear polarizer in front of detectors. All measurements were done in atmosphere and at room temperature.

**Acknowledgments**

We acknowledge Prof. Yuen-Ron Shen and Prof. Wenzhong Bao for discussions. The work at Fudan University was supported by the National Basic Research Program of China (Grant Nos. 2014CB921601, 2016YFA0301002), National Natural Science Foundation of China (Grant No. 91421108, 11622429, 11374065), and the Science and Technology Commission of Shanghai Municipality (Grant No. 16JC1400401).

**Author contributions**

**Competing interests**

The authors declare that they have no competing interests.

**Figures**

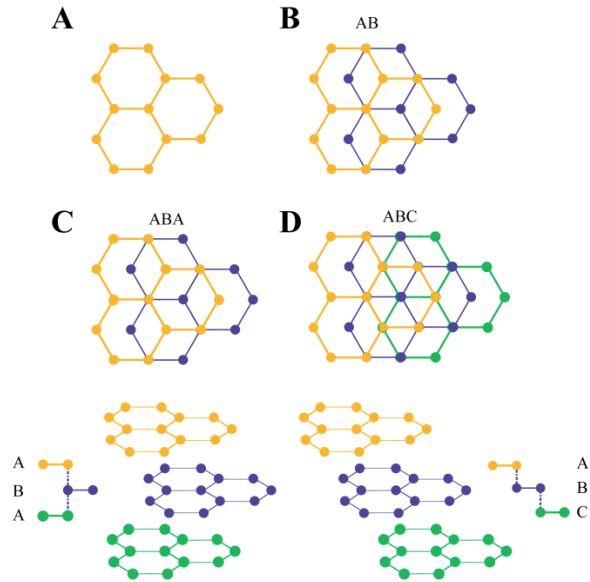

**Fig. 1. Stacking orders in few-layer graphene.** **(A)** graphene monolayer. **(B)** AB stacked graphene bilayer. **(C)**, **(D)** two most common polymorphs of graphene trilayer, ABA (Bernal) and ABC (rhombohedral), respectively. The top, middle, and bottom layers are labeled in yellow, purple, and green, respectively. In ABA stacking, the top layer lies exactly on top of the bottom layer; in ABC stacking, one sublattice of the upper layer lies above the center of the hexagons in the lower layer.

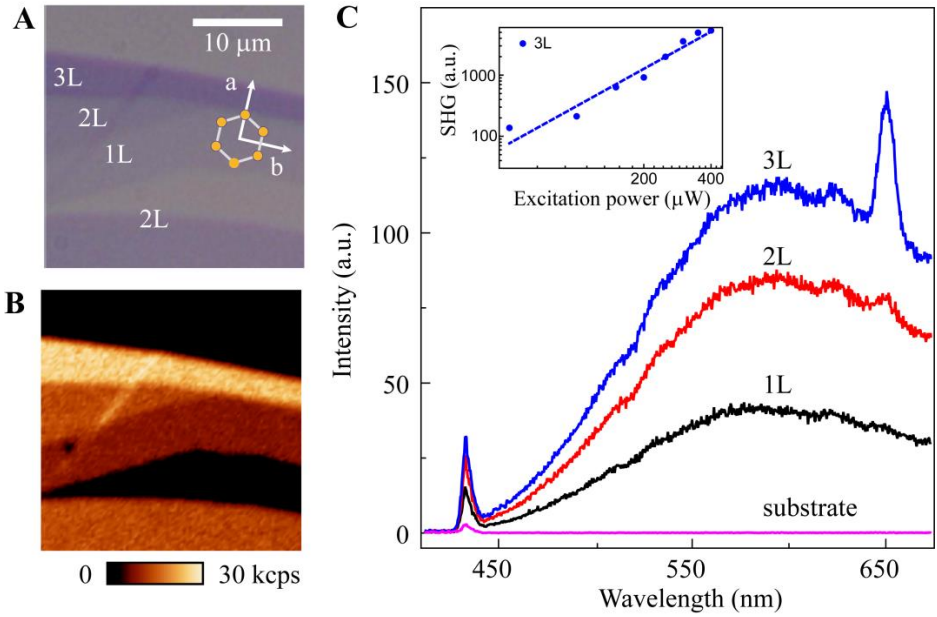

**Fig. 2. Optical microscopy and spectroscopy of few-layer graphene.** (**A**) Bright-field optical microscopy of mechanically exfoliated few-layer graphene on silicon wafer with 300 nm thick $SiO_2$. The crystalline orientation was overlaid on the image, as determined from the azimuthal polarization patterns in Figure 3. (**B**) Up-converted nonlinear photoluminescence microscopy of the same area as (**A**). The sample was excited by femtosecond pulses at wavelength of 1300 nm and average power of 0.5 mW without any damages. The collected signal was spectrally from 425 nm to 675 nm. (**C**) Corresponding up-converted optical spectra from monolayer (1L), bilayer (2L), trilayer (3L) and bare substrate, respectively. Besides the broad spectra from nonlinear photoluminescence, third harmonic generation (THG) at 433 nm and second harmonic generation (SHG) at 650 nm were also observed. The inset of (**C**) plotted the power dependence of SHG, after subtracting the nonlinear photoluminescence background, on trilayer graphene in a log-log scale. The dotted data were fitted linearly with a slope of 2.02 ± 0.18.

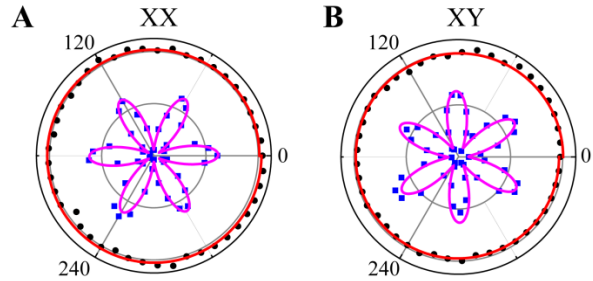

**Fig. 3. Azimuthal polarization patterns of SHG and up-converted nonlinear photoluminescence on ABA trilayer. (A), (B)** The polarizations of the excitation and signal beams are set in parallel (XX) and perpendicular (XY), respectively. The nonlinear photoluminescence and SHG signals were shown in black and blue dots, respectively. The intensity of nonlinear photoluminescence was integrated from 640 nm to 660 nm and reduced by 10 times for clarity. While the nonlinear photoluminescence was fitted with a constant azimuthally, the SHG signals were fitted to functions of $A\cos^2(3\theta)$ and $A\sin^2(3\theta)$ for (a) and (b), respectively.

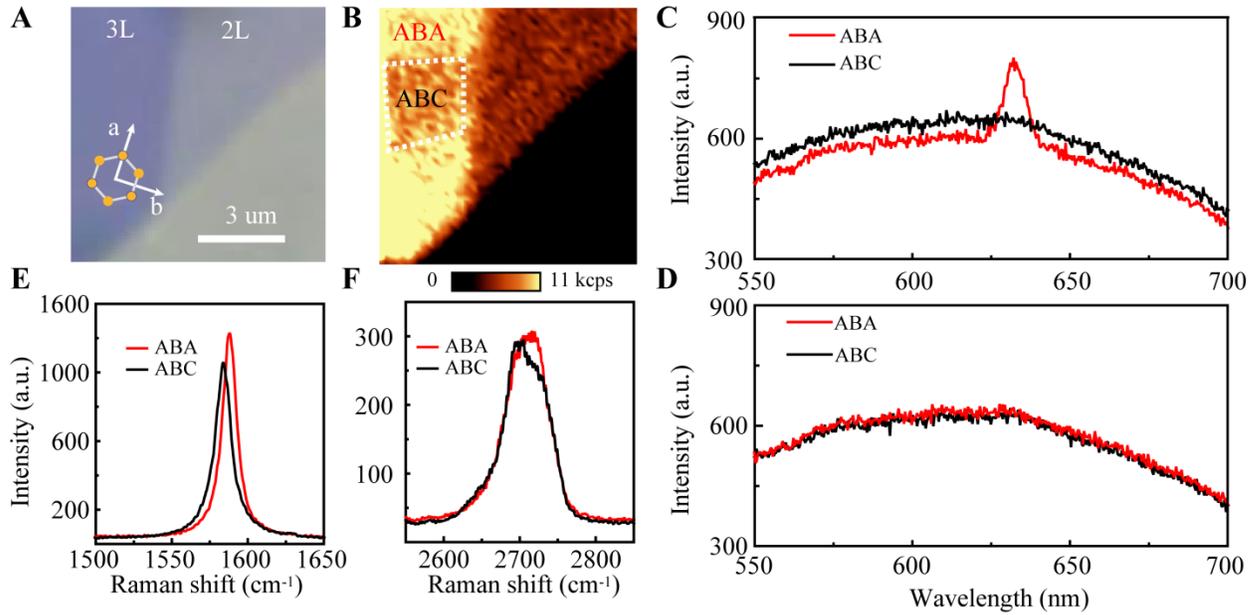

**Fig. 4. Imaging the stacking order of ABA and ABC trilayers.** (**A**) Bright-field optical microscopy image of few-layer graphene with trilayer on the left and bilayer on the right. The crystalline orientation was also overlaid on the image. (**B**) Corresponding nonlinear microscopy of the same area as (**A**) with excitation wavelength of 1266 nm. The spectral bandpass range for nonlinear signal was centered at 633 nm and narrowed to 6 nm, so that SHG could be more pronounced over NPL. The dwell time at each pixel was 25 ms. The dimmer region from trilayer graphene in (**B**) is enclosed by dotted lines and attributed to ABC stacking. (**C**), (**D**) Nonlinear optical spectra from ABA and ABC trilayer regions, marked in (**B**), when both the excitation and signal beams were linearly polarized along the arm chair or zigzag directions, respectively. (**E**), (**F**) Corresponding G and 2D Raman peaks from ABA (red) and ABC (black) trilayers, respectively. The spectra were excited at the wavelength of 532 nm and the power of 1 mW.

# Supplementary Materials for

# "Stacking-symmetry governed second harmonic generation in graphene trilayers"

Yuwei Shan,[1] Yingguo Li,[1] Di Huang,[1] Qingjun Tong,[2] Wang Yao,[2] Wei-Tao Liu,[1,3] Shiwei Wu[1,3*]

[1] State Key Laboratory of Surface Physics, Key Laboratory of Micro and Nano Photonic Structures (MOE), and Department of Physics, Fudan University, Shanghai 200433, China.

[2] Department of Physics and Center of Theoretical and Computational Physics, University of Hong Kong, Hong Kong, China.

[3] Collaborative Innovation Center of Advanced Microstructures, Nanjing 210093, China.

[*] Corresponding email: swwu@fudan.edu.cn

**The supplementary materials include:**

    **Supplementary Figures S1-S4**

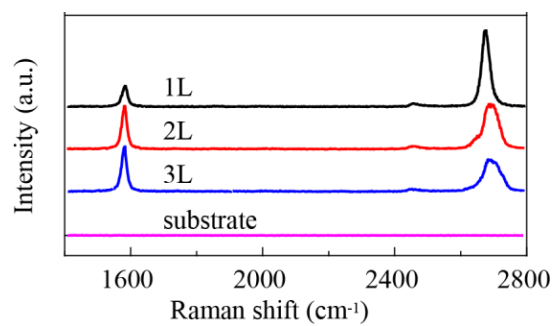

**Supplementary Figure S1.** Raman spectra of the sample from monolayer (1L), bilayer (2L) and trilayer (3L), along with the background from the bare substrate. The excitation wavelength was 532 nm. For the thinnest layer, the 2D peak (~2676 cm$^{-1}$) has a single Lorentzian line shape indicating a monolayer. For thicker layers, the 2D bands are broadened as composed of multiple peaks, and intensity of the G band (~1588 cm$^{-1}$) is roughly proportional to the number of layers.

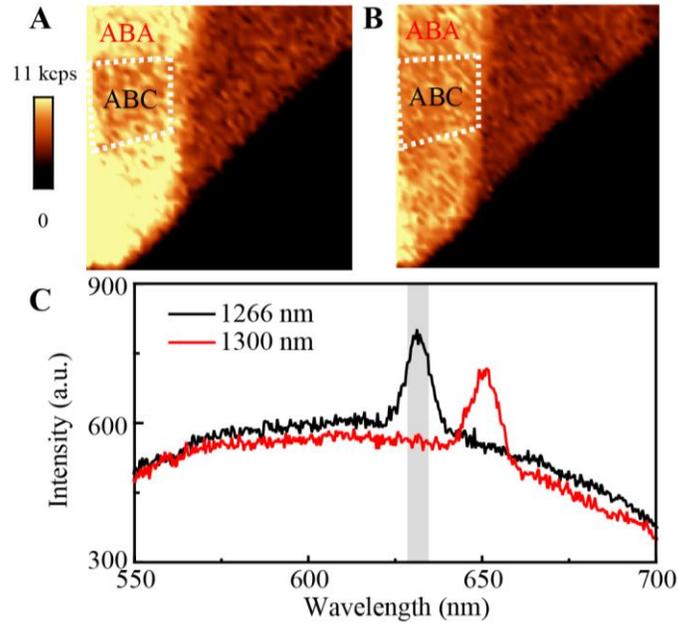

**Supplementary Figure S2.** Comparison of nonlinear microscopy at different excitation wavelength. **(A), (B)** Nonlinear microscopy of the graphene trilayer in Figure 4 with excitation wavelength of 1266 nm and 1300 nm, respectively. The dwell time at each pixel was 25 ms. (**C**) Corresponding nonlinear optical spectra from the ABA trilayer region. Both the excitation and signal beams were linearly polarized along the arm chair direction. The spectral bandpass range for nonlinear signal was centered at 633 nm and narrowed to 6 nm, which is grayed in the spectra.

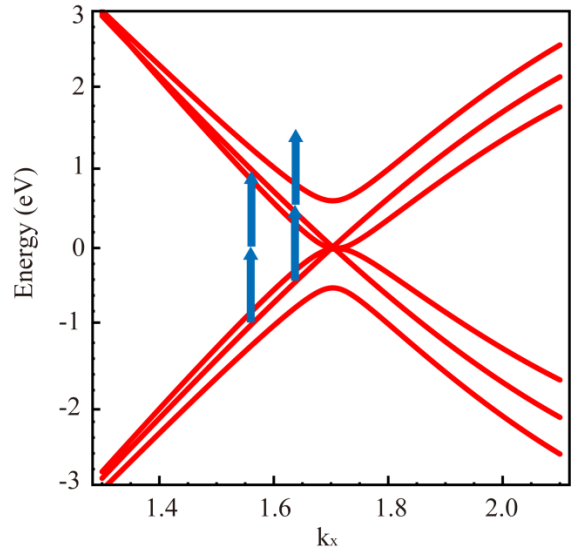

**Supplementary Figure S3.** Schematic showing the resonant transitions in the SHG process. The one-photon or two-photon energy (blue arrows) could match with the interband transition in the band structure of ABA graphene trilayer (red curves). The resonance effects lead to the strong SHG we observed.

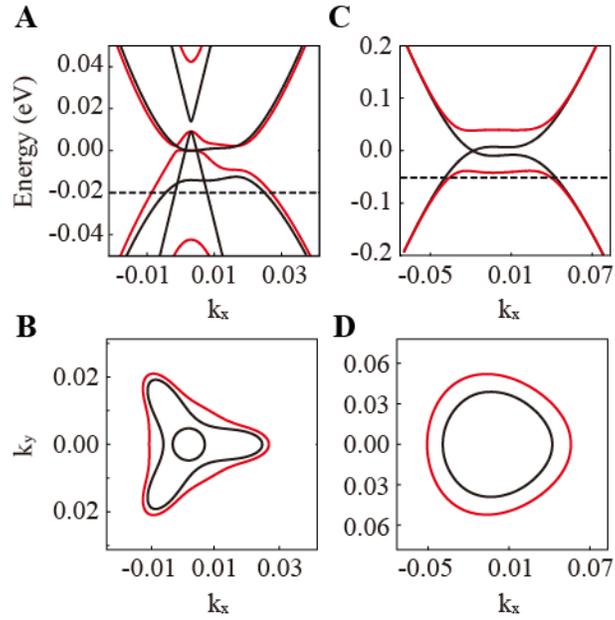

**Supplementary Figure S4.** Stacking dependent properties in ABA and ABC trilayers. **(A)**, **(B)** Calculated electronic band structure and contour of ABA trilayer, respectively. **(C)**, **(D)** Calculated electronic band structure and contour of ABC trilayer, respectively. The calculation followed the low-energy effective models (*36, 37*). The black and red curves correspond to the band structures without ($U_1 = 0$ eV, $U_2 = 0$ eV, $U_3 = 0$ eV) and with ($U_1 = 0.04$ eV, $U_2 = 0$ eV, $U_3 = -0.04$ eV) a perpendicular electric field through the trilayer graphene, respectively. The contours in **(B)** and **(D)** are taken from the dotted line in **(A)** and **(C)** at the energy level of -0.02 eV and -0.05 eV, respectively. The effect of trigonal warping in ABA trilayer is stronger than that in ABC trilayer.